\documentclass[pmlr,twocolumn,10pt]{jmlr} 

\usepackage{float}
\usepackage{booktabs}
\usepackage{siunitx}
\usepackage[switch]{lineno}
\usepackage{multirow}

\theorembodyfont{\upshape}
\theoremheaderfont{\scshape}
\theorempostheader{:}
\theoremsep{\newline}

\jmlrvolume{LEAVE UNSET}
\jmlryear{2024}
\jmlrworkshop{Machine Learning for Health (ML4H) 2024} 

\title{Fluoroformer: Scaling multiple instance learning to multiplexed images via attention-based channel fusion}

\author{
    Marc Harary \Email{marc@ds.dfci.harvard.edu} \\ 
    \addr  Dana-Farber Cancer Institute, Boston, MA, USA
    \AND
    Eliezer M. Van Allen \Email{EliezerM\_VanAllen@dfci.harvard.edu} \\ 
    \addr  Dana-Farber Cancer Institute, Boston, MA, USA
    \AND
    William Lotter \Email{lotterb@ds.dfci.harvard.edu} \\
    \addr  Dana-Farber Cancer Institute, Boston, MA, USA
}
\begin{document}

\maketitle

\begin{abstract}
    Though multiple instance learning (MIL) has been a foundational strategy in computational pathology for processing whole slide images (WSIs), current approaches are designed for traditional hematoxylin and eosin (H\&E) slides rather than emerging multiplexed technologies. Here, we present an MIL strategy, the Fluoroformer module, that is specifically tailored to multiplexed WSIs by leveraging scaled dot-product attention (SDPA) to interpretably fuse information across disparate channels. On a cohort of 434 non-small cell lung cancer (NSCLC) samples, we show that the Fluoroformer both obtains strong prognostic performance and recapitulates immuno-oncological hallmarks of NSCLC. Our technique thereby provides a path for adapting state-of-the-art AI techniques to emerging spatial biology assays.
\end{abstract}

\paragraph*{Data and Code Availability}
The ImmunoProfile dataset \citep{lindsay2023immunoprofile} used in this manuscript has not been IRB approved for public release. Code is available at \url{https://github.com/lotterlab/fluoroformer}.

\paragraph*{Institutional Review Board (IRB)}
All patients in the ImmunoProfile cohort provided consent under an institutional research protocol (DFCI 11-104, 17-000, 20-000).

\section{Introduction}
Multiple instance learning (MIL) has emerged as the \textit{de facto} standard approach in computational pathology for generating predictions from whole slide images (WSIs) \citep{ilse2018attention, maron1997framework, carbonneau2018multiple, lu2021data}. The typical MIL pipeline consists of 1) dividing the WSI into smaller image patches, 2) extracting lower dimensional embeddings for each patch from a pre-trained neural network, 3) pooling embeddings across patches to create a slide-level summary vector, and 4) generating slide-level predictions for the particular task at hand. Compared to traditional strategies such as training patch-level predictors that rely exclusively on clinician-annotated regions of interest (ROIs), MIL enables weakly-supervised training on entire WSIs, thereby offering enhanced scalability, reduced sampling bias, and potentially superior performance \citep{zhou2018brief}.

\begin{figure*}[ht!]
\includegraphics[width=\linewidth]{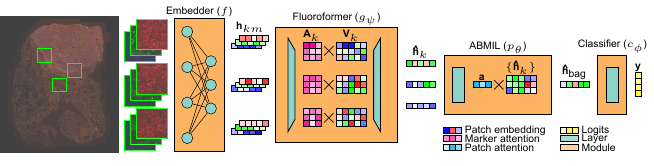}
\caption{Overview of the Fluoroformer strategy for multiplexed imaging. Mathematical symbols are defined in text.}
\label{fig1}
\end{figure*}

Thus far in computational pathology, MIL pipelines have largely been confined to traditional hematoxylin \& eosin (H\&E)-stained WSIs \citep{wilson2021challenges, ghahremani2022deep}. While H\&E staining can provide detailed morphological information, it fails to explicitly capture important proteins and other complex biomarkers that indicate cell phenotype and state \citep{lee2020multiplex, peng2023multiplex, munoz2022cyclic}. 
In contrast, emergent techniques in spatial biology such as multiplex immunofluorescence (mIF) enable the imaging of many biomarkers simultaneously in tissue samples while preserving spatial context (Figure~\ref{fig1}). These techniques result in rich, multi-channel ($\sim$5-50) images that have advanced our understanding of diseases ranging from neurodegenerative disorders \citep{munoz2022cyclic} to cancer \citep{lee2020multiplex, peng2023multiplex}. Conversely, mIF images are often analyzed using hand-engineered features, such as the counts of discrete biomarkers within clinician-defined ROIs \citep{wilson2021challenges}. More recent efforts have pointed to the potential of using deep learning to improve performance on downstream tasks, but these efforts have also focused on ROIs rather than expanding to WSIs \citep{hoebel2024deep, sorin2023single, wu2022graph}. There is therefore a pressing need to optimize MIL methods for mIF in order to yield the benefits of both weakly-supervised training and the rich information provided by spatial assays. Doing so, however, presents several challenges. The disparate channels must be somehow combined, and, moreover, ideally would be done so flexibly given that the number of channels can vary between mIF protocols.

Here, we present the Fluoroformer, a Transformer-like neural network module designed to interpretably scale MIL to multiplex images. Leveraging scaled dot-product attention (SDPA) \citep{vaswani2017attention}, it fuses the information from disparate multiplexed channels into a single summary vector for each patch, enabling the subsequent pooling of the patch embeddings via standard attention-based MIL (ABMIL) mechanisms. Importantly, the Fluoroformer produces attention matrices for each patch that may offer insights into cell-cell interactions and biological structures. Using a cohort of 434 non-small cell lung cancer (NSCLC) samples and their corresponding mIF WSIs, we find that the Fluoroformer demonstrates strong performance in predicting patient prognosis. Analysis of the channel-wise attention matrices offers insights into immune-tumor interactions that potentially associate with prognosis. Our approach therefore bridges spatial biology techniques with state-of-the-art artificial intelligence approaches to maximize the potential of this emerging field. 

\begin{figure*}[t!]
\includegraphics[width=\linewidth]{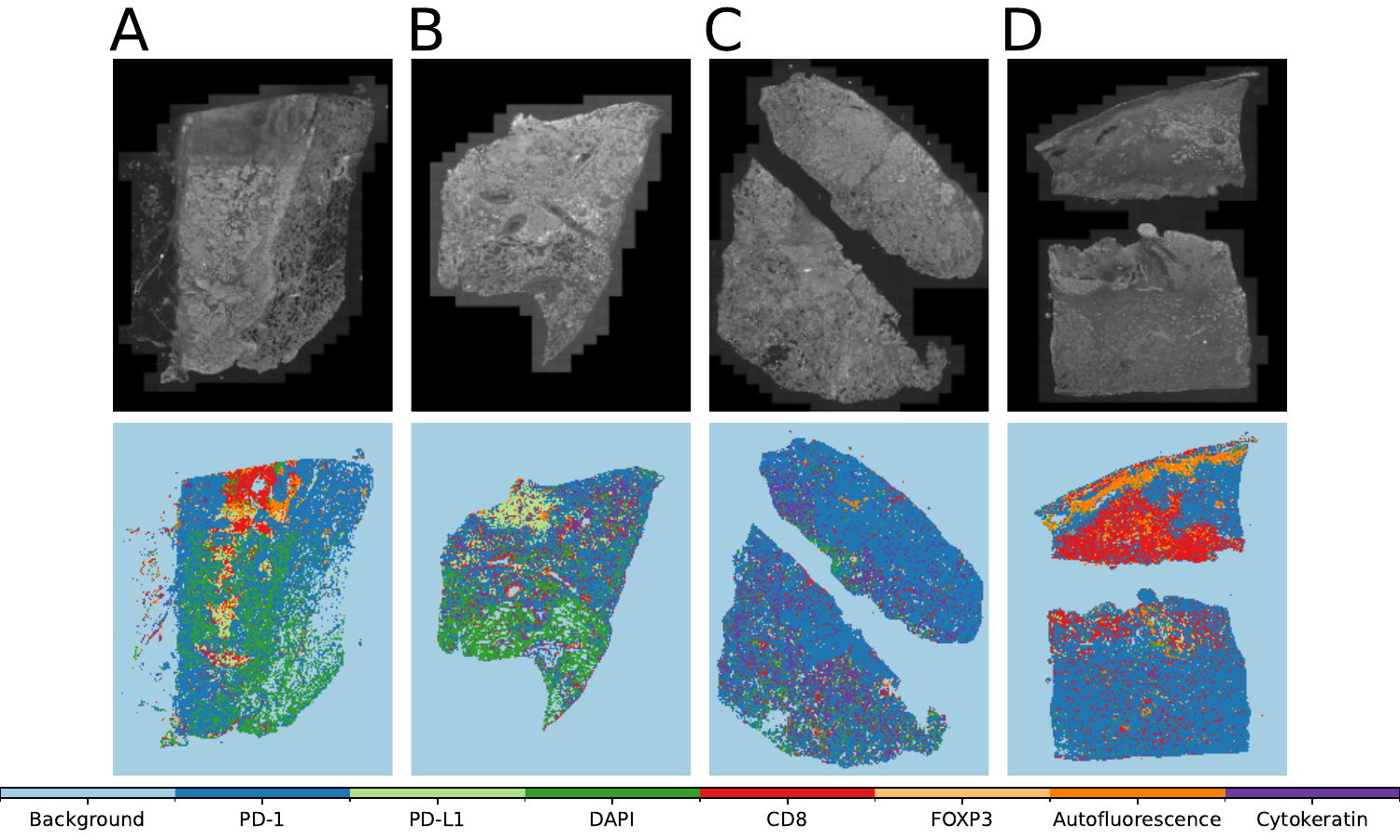}
\caption{Heatmaps of the most highly attended markers in each patch of selected mIF images for the Fluoroformer model with a ResNet50 embedder. Patterns are observed such as DAPI being a highly attended marker for alveolar tissue (A, B) and CD8-attended regions appearing at tumor margins and sporadically within the tumor (A, C, D).}
\label{fig2}
\end{figure*}

\section{Related work}

\subsection{Attention-based fusion strategies in histopathology}
Attention-based pooling has emerged as among the most popular information aggregation strategies for H\&E histopathology slides. 
This includes attention-based multiple instance learning (ABMIL), a benchmark MIL approach for H\&E WSIs that consists of a traditional gated attention mechanism \citep{ilse2018attention} as further described below. Several works have also applied scaled dot-product attention (SDPA) and multihead attention (MHA) \citep{vaswani2017attention} to H\&E WSIs. A prime example is TransMIL \citep{shao2021transmil}, which leverages MHA rather than gated attention to pool embeddings across the full slide. Beyond aggregating information across patches, attention mechanisms have also been used to fuse multiple modalities. Recently, MCAT \citep{chen2021multimodal} has leveraged SDPA to integrate genomic data and H\&E embeddings. 

\subsection{Deep learning for multiplexed pathology images}
Several recent studies have applied deep learning to mIF. For instance, \cite{sorin2023single} applied a ResNet-based \citep{he2016deep} pipeline to the ROIs in NSCLC samples and found improved prognostic performance compared to models relying on traditional features. \cite{wu2022graph} developed a graph neural network approach based on point patterns of cell phenotypes, also for mIF ROIs, and likewise observed higher prognostic performance relative to hand-engineered metrics. While these works demonstrate the promise of deep learning applied to multiplexed images, they require human input to define relevant regions, limiting scalability and underutilizing the full amount of data present in WSIs.

\section{Methodology}

Our Fluoroformer approach adapts MIL to multiplex WSIs via an attention-based channel fusion mechanism, inserted as an encapsulated module between the typical patch embedding and slide-level aggregation stages. We first provide a preliminary summary of MIL and ABMIL before describing our optimizations for multiplexing.

\begin{figure}[t!]
\includegraphics[width=\linewidth]{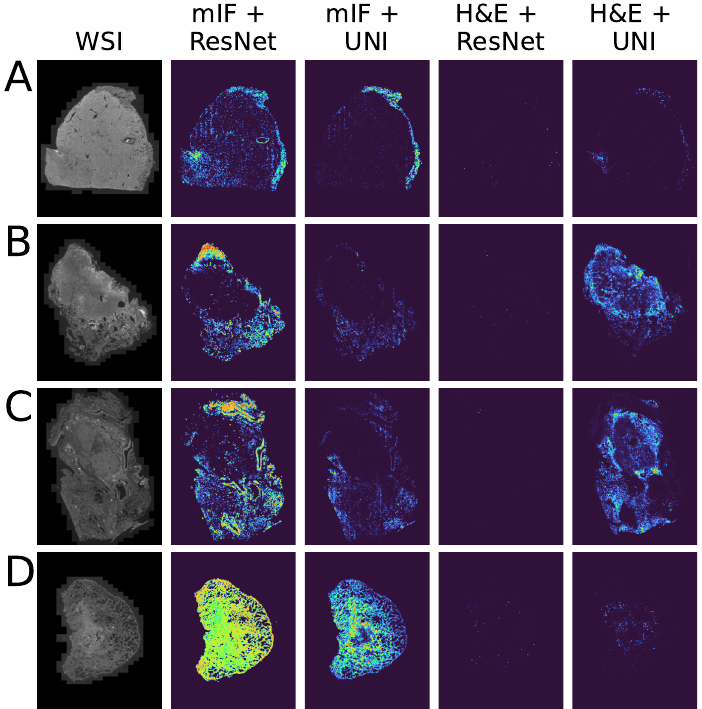}
\caption{Heatmaps of patch attention for each combination of imaging modality and neural embedder. The mIF heatmaps often demonstrate higher spatial continuity and are concentrated towards regions adjacent to the tumor mass (e.g., B, C), whereas the H\&E maps are often sparser and concentrate on inner tumor regions.}
\label{fig3}
\end{figure}

\subsection{Preliminaries: MIL and ABMIL}

Multiple instance learning (MIL) \citep{ilse2018attention, maron1997framework, carbonneau2018multiple} is typically formulated as follows. Each sample consists of a set-based data structure $\mathbf X = \left\{ \mathbf x_k \right\}$ known as ``bag'' that contains, in a permutation-invariant fashion, $K$ separate ``instances'' $\mathbf x_k$, where each instance $\mathbf x_k \in \mathcal X$ and the bag $\mathbf X \in \mathcal X^*$. The task is weakly supervised \citep{zhou2018brief}, meaning that $\mathbf X$ is associated with a global label $\mathbf y \in \mathcal Y$.

MIL models consist of a pooling operation $p$ that aggregates all instances, paired with a learnable classifier $c_\phi$ \citep{ilse2018attention, maron1997framework, carbonneau2018multiple}. For biomedical image processing, in which images may often be exceptionally large, instances consist of small patches of a larger image \citep{sudharshan2019multiple, javed2022additive, chen2024towards, xu2024whole}. These are embedded via a ``featurizer'' $f: \mathcal X \to \mathcal H$ into a latent space $\mathcal H = \mathbb R^{d_\text{emb}}$ to produce an embedded bag $H \in \mathcal H^K$:
\begin{equation}
    H = \left\{ \mathbf h_k = f \left(\mathbf x_k \right) \mid \mathbf x_k \in \mathbf X \right\}.
\end{equation}
Popular choices for $f$ are ResNet50 \citep{he2016deep} trained on ImageNet or, more recently, specialized foundation models like UNI \citep{chen2024towards} and Gigapath \citep{xu2024whole} that have been trained in a self-supervised fashion on large domain-specific (H\&E) datasets. Notably, $f$ is frozen prior to training of the pooling and classification operations in standard approaches.

In ABMIL, the pooling operation consists of a learnable attention module $p_\theta: \mathcal H^K \to \mathcal H$ that computes a weighted sum of the embedded bag \citep{ilse2018attention}. Most popularly, including in CLAM \citep{lu2021data}, a single- or double-gated attention mechanism computes a weighted sum of the patch vectors via an attention vector $\mathbf a \in \mathbb R^K$ \citep{ilse2018attention}:
\begin{align}
    \mathbf h_{\text{bag}} = p_\theta \left( H \right) = \sum_{k} a_k \mathbf h_k.
\end{align}
In the double-gated variant, we have
\begin{equation}
    a_k = \text{softmax}\left( \mathbf w^\top \left( \text{tanh} \left( \mathbf V \mathbf h_k^\top \right) \odot \text{sigm} \left( \mathbf U \mathbf h_k^\top \right) \right) \right),
\end{equation}
where $\mathbf w \in \mathbb R^{d_\text{emb} \times 1}$, $\mathbf V \in \mathbb R^{d_\text{emb} \times d_\text{emb}}$, and $\mathbf U \in \mathbb R^{d_\text{emb} \times d_\text{emb}}$ are learned parameters such that $\theta = \left(\mathbf w, \mathbf V, \mathbf U \right)$; $\odot$ denotes the Hadamard product; and $\text{sigm}$ the sigmoid non-linearity. Biases are implied.

The final module, $c_\phi$, maps the aggregated embeddings $\mathbf h_\text{bag}$ to the output space $\mathcal Y$. Typically, both in popular implementations and our own below, $c_\phi$ consists of a single linear layer that returns the output logits
\begin{equation}
    \mathbf y = c_\phi(\mathbf h_{\text{bag}}) = \mathbf w^\top  \mathbf h_{\text{bag}},
\end{equation}
where $\phi = \mathbf w$ (again omitting the bias term).

\subsection{Fluoroformer: Leveraging MIL for multiplexing}

Multiplexed imaging adds another dimension along which aggregation must be performed, namely the large number ($\sim$5-50) of disparate channels corresponding to separate biomarkers. Rather than trying to convert all of these channels into an RGB image or developing an embedder that can directly process many channels, we apply a pre-trained featurizer to each channel separately. To do so, each channel is first duplicated thrice along the RGB channel-dimension to produce a gray-scale image that can be processed by standard featurizers. Given $M$ channels, the global bag corresponding to a full sample now consists of
\begin{equation}
    \mathbf X = \left\{ \mathbf x_{km} \right\}, \quad \text{where} \quad \mathbf x_{km} \in \mathcal X^m,
\end{equation}
or, equivalently,
\begin{equation}
    \mathbf H = \left\{ \mathbf h_{km} \right\}, \quad \text{where} \quad \mathbf h_{km} \in \mathcal H^m.
\end{equation}


To leverage the benefits of multiplexing in combining features across channels, we borrow insights from natural language processing (NLP). The semantic information contained in a sentence is not determined by each of its constituent tokens independently; rather, the tokens interact in pairwise dependencies to collectively produce the semantic significance of the full sequence. Likewise, multiplexing is often employed to capture complex relationships between each channel (e.g., tumor and immune cell interactions), rather than to simply image several features simultaneously \textit{per se}.

We therefore propose fusing embeddings across channels in each patch using SDPA, mirroring the message-passing between tokens that is performed by Transformers.
A secondary advantage of such a pairwise attention mechanism lies in explainability; meaningful relationships between individual channels can be captured in the attention matrix of each patch, potentially identifying novel patterns while also reflecting the complex structure in the corresponding region of the WSI.

\subsubsection{Marker attention}
Mathematically, the SDPA operation in the Fluoroformer architecture thereby fills the role of a fourth, learnable operation $g_\psi: \mathcal H^m \to \mathcal H$ to preliminarily pool along the channel dimension:
\begin{equation}
    \mathbf{\hat h}_k = g_\psi \left( \left\{ \mathbf h_{km} \right\} \right).
\end{equation}
Treating each $k$th patch as a ``sentence'' consisting of $M$ tokens in an $d_{\text{emb}}$-dimensional latent space \citep{otter2020survey}, we compute ``query,'' ``key,'' and ``value'' embeddings $\mathbf Q_k, \mathbf K_k, \mathbf V_k \in \mathbb R^{M \times d_\text{hid}}$ via standard linear layers \citep{vaswani2017attention}, then employ the following formula:
\begin{align}
    \mathbf a_{km} &= \mathbf A_{km} \mathbf V_{km} =\text{softmax} \left( \frac{\mathbf Q_{km} \mathbf K_{km}^\top}{\sqrt{d_\text{hid}}} \right) \mathbf V_{km},
\end{align}
where $d_\text{hid}$ is a hidden dimension and $\mathbf A_{km}$ is the pairwise attention matrix. 

Because we will almost always have $d_\text{emb} \gg M$, we minimize computational overhead by adding a preliminary bottleneck that contracts the embedding dimension $d_\text{emb}$ to a hidden dimension $d_\text{hid}$. 
We let $\mathbf{ \tilde h}_{km}$ denote the contracted embedding.

\subsubsection{Marker normalization}

In keeping with the standard Transformer architecture \citep{vaswani2017attention}, we then employ two skip connections \citep{he2016deep} each followed by ``marker normalization'' layers. Analogous to layer normalization \citep{ba2016layer} in standard NLP models, these have the motivation of both stabilizing training and ensuring that no one channel dominates the patch when performing mean pooling. Specifically, for each $k$th patch and $m$th channel, the sum $\mathbf a_{km}$ is updated with its residual
\begin{equation}
    \mathbf a_{km} \gets \mathbf a_{km} + \mathbf{\tilde h}_{km}.
\end{equation}
The resulting tensor is normalized by computing the statistics
\begin{align}
    \mu^{\text{(SDPA)}}_{km} &= \frac{1}{d_{\text{hid}}} \sum_{i=1}^{d_{\text{hid}}} a_{kmi} \\
    \sigma^{\text{(SDPA)}}_{km} &= \sqrt{\frac{1}{d_{\text{hid}}} \sum_{i=1}^{d_{\text{hid}}} \left( a_{kmi} - \mu^{\text{(SDPA)}}_{km} \right)^2} + \varepsilon,
\end{align}
where $\varepsilon$ is a small constant to prevent division by 0 \citep{ba2016layer}. The layer then updates each channel via
\begin{align}
    \mathbf a_{km} \gets \frac{\mathbf a_{km} - \mu_{km}^{\text{(SDPA)}}}{\sigma_{km}^{\text{(SDPA)}}} \boldsymbol \gamma^{\text{(SDPA)}} + \boldsymbol \beta^{\text{(SDPA)}},
\end{align}
where $\boldsymbol \gamma^{\text{(SDPA)}}$ and $\boldsymbol \beta^{\text{(SDPA)}}$ are learnable affine parameters. 

Next, the bottleneck is inverted by a simple linear layer. To prevent loss of information, a second skip connection adds the original quantity $\mathbf h_{km}$ back to $\mathbf a_{km}$ followed by another round of patch normalization with corresponding quantities $\boldsymbol \mu_{km}^{\text{(bottleneck)}}$ and $\boldsymbol \sigma_{km}^{\text{(bottleneck)}}$. GELU is used following each linear transform \citep{hendrycks2016gaussian}.

Finally, mean pooling is performed along the marker dimension, creating a summary $\mathbf{\hat h}_k$ of the markers and their interactions within the $k$th patch:
\begin{equation}
    \mathbf{\hat h}_k = \frac{1}{M} \sum_{m=1}^M \mathbf a_{km}.
\end{equation}
Having effectively eliminated the additional dimension, the pooled bag $\left\{ \mathbf{\hat h}_k \right\}$ becomes equivalent to a non-multiplexed input such that any MIL aggregation strategy can be applied thereafter. In our case, we pass the fused bag to a standard ABMIL double-gated attention module $p_\theta$ and linear classifier $c_\phi$ as described above.


\section{Experimental Details}

We trained the Fluoroformer model to perform survival prediction for non-small cell lung cancer (NSCLC) WSIs. For each sample in the utilized cohort, both mIF and H\&E pathology slides were available, allowing us to directly compare the performance of the Fluoroformer to state-of-the-art H\&E ABMIL approaches. For both, we consider two patch embedders, namely ResNet50 \citep{he2016deep} pre-trained on ImageNet and UNI \citep{chen2024towards}, a histopathology foundation model consisting of a vision transformer (ViT) \citep{dosovitskiy2020image} trained via the Dinov2 algorithm \citep{oquab2023dinov2} on a large H\&E dataset. As an additional baseline, we compare to a Cox proportional hazards (CoxPH) model \citep{cox1972regression} fit using the intratumoral cell densities of the mIF biomarkers, as described below. 

\subsection{Dataset}
The NSCLC dataset used consists of 434 primary-site tumor samples from 414 patients and resulted from the ImmunoProfile project \citep{lindsay2023immunoprofile}, a prospective mIF effort performed at the Dana-Farber Cancer Institute from 2018-2022. The ImmunoProfile assay stains for four immune markers (CD8, FOXP3, PD-L1, PD-1), cytokeratin (Cyto) as a tumor marker, and DAPI as a counterstain for nucleus detection. Briefly, CD8 is a marker for cytotoxic T cells that can attack tumor cells \citep{raskov2021}. FOXP3 is a marker for regulatory T cells that can indicate immune suppression \citep{Rudensky2011}. PD-1 and PD-L1 are involved in immune inhibition and can be expressed by both immune and tumor cells \citep{han2020pd}. Along with these biomarkers, an autofluorescence channel is included in each mIF WSI in the dataset. 
The intratumoral density of cells positive for each immune marker (CD8, FOXP3, PD-L1, and PD-1) and the PD-L1 tumor proportion score (TPS) have also been calculated for each sample based on expert-annotated ROIs.
These cell density metrics are commonly used in mIF studies \citep{lindsay2023immunoprofile}, where PD-L1 TPS quantifies the percent of tumor cells that are positive for PD-L1.



For each sample, the H\&E and mIF WSIs were acquired from different tissue sections of the same tumor sample and were not registered. Follow-up time and survival status (deceased or censored) were also recorded, meaning that each sample consists of the tuple $\left(\mathbf{X}_{\text{H\&E}}^{(i)}, \mathbf{X}_{\text{mIF}}^{(i)}, t^{(i)}, c^{(i)} \right)$, where $t^{(i)} \in \mathbb R$, and $c^{(i)} \in \left\{ 0, 1 \right\}$. As an unselected clinical population, the dataset consists of tumors across different stages (306 or 70.5\% low-stage, 125 or 28.8\% high stage, 3 or 0.7\% unknown stage) and different treatment regimes (97 or 22.4\% receiving immunotherapy, 334 or 77.0\% receiving treatment other than immunotherapies, 3 or 0.7\% with unknown treatment), representative of a real-world clinical cohort. All data used in the experiments are de-identified. 


\subsection{Preprocessing}

For the H\&E images, preprocessing consisted of first identifying foreground patches and then embedding each patch using a pre-trained embedder. For a given WSI, foreground patches were identified by applying Otsu’s algorithm \citep{otsu1975threshold} to a grayscaled version of the WSI after downsampling by 224. The original RGB image patches corresponding to the foreground were then used as input to the embedder to create an embedding ${\mathbf h}_k\in \mathbb R^{d_\text{emb}}$ for each patch. We perform experiments with two different embedders: UNI \citep{chen2024towards} ($d_\text{emb} = 1024$) and ResNet50 \citep{he2016deep} (pre-trained on ImageNet; $d_\text{emb} = 2048$).

For identification of foreground patches for mIF, all seven gray-scale WSI channels were downsampled and thresholded separately, which was followed by a pixel-wise OR operation across each of the seven binary masks to pool along the channel dimension. For each foreground patch, each channel in the patch was then repeated three times along an added color dimension to create a gray RGB image patch and embedded channel-wise to produce a matrix $\mathbf h_{k} \in \mathbb R^{M \times d_\text{emb}}$ for each $k$th patch.

\subsection{Task and objective function}

Pursuing a common strategy for prognostication in deep learning, we train the Fluoroformer and standard ABMIL baseline models to regress the discrete hazard and survival functions \citep{cox1972regression, zadeh2020bias}, given by
\begin{align}
    h \left( t \mid \mathbf X \right) &:= P(T=t \mid T\geq t, \mathbf X)
\end{align}
and
\begin{align}
    S \left( t \mid \mathbf X \right) &:= P(T>t \mid \mathbf X) \nonumber \\
    &= \prod_{s=1}^t \left(1 - h \left(s \mid \mathbf X \right) \right).
\end{align}

When the hazard function is discretized \citep{katzman2018deepsurv, zadeh2020bias}, the hazard ratio is predicted for $N_\text{bin}$ intervals defined by cutoffs $\left\{-\infty, t_1, t_2, \ldots, t_{N_\text{bin} - 1} \right\}$. We use four bins based on quartiles of event times in the dataset. The output label for a network is correspondingly a logit vector $\mathbf{\hat y} \in [0, 1]^{N_\text{bin}}$ such that $\hat y_i$ is equal to the probability of the event occurring in the interval $\left[ t_{i-1}, t_{i} \right]$.
As proposed by \citep{zadeh2020bias}, we then use the log-likelihood objective given by
\begin{align}
    \mathcal L \left(\mathbf X, t, c \right) = & -c \log S \left( t \mid \mathbf X \right) \nonumber \\
    &- \left(1 - c \right) \log S \left( t - 1 \mid \mathbf X \right) \nonumber \\
    &-  \left(1 - c \right) \log h \left( t \mid \mathbf X \right).
\end{align}

\subsection{Performance metrics}

To evaluate the performance of each model, five-fold cross-validation with stratification by patient was performed. Each split involved using three folds for training, a fourth for validation, and the fifth for testing. For each test fold and each sample therein, a risk score was computed via
\begin{equation}
    r^{(i)} := \sum_{j=1}^{N_\text{bin}} S^{(i)}_j,
\end{equation}
where $S^{(i)}_j$ is the survival rate for the $j$th bin. 
The concordance index (C-index) was then computed between the risk scores and the observed outcomes ($t^{(i)}, c^{(i)}$), where the C-index is a standard metric in survival analysis and represents the probability of correctly ranking pairs of samples. As is standard in histopathology prognostication tasks, such as benchmarks using TCGA, the models are trained solely based on the WSIs and do not receive other patient or tumor metadata as input (e.g., treatment, age).
We additionally compute the C-index for a CoxPH model using the intratumoral cell densities (CD8, FOXP3, PD-L1, PD-1) and PD-L1 TPS as covariates.
The same cross validation folds are used for fitting and testing this baseline model as the MIL models. 


\subsection{Attention heatmap metrics}

We assessed the patch-wise attention vectors $\left( \mathbf a \right)$ produced by the models both qualitatively and quantitatively. For the latter, we consider the spatial autocorrelation of the output attention heatmaps with the following intuition: Neighboring patches in tissue samples often contain similar features, meaning that robust heatmaps should likely exhibit smoother spatial variation (i.e., higher spatial autocorrelation). Mathematically, spatial autocorrelation can be quantified using Moran's I (MI) \citep{moran1950notes}, which ranges from -1 (perfect negative correlation) to 1 (perfect positive correlation). The formula employed was
\begin{equation}
I := \frac{N \sum_{i=1}^{N} \sum_{j=1}^{N} w_{ij} (x_i - \bar{x})(x_j - \bar{x})}{\left(\sum_{i=1}^{N} \sum_{j=1}^{N} w_{ij}\right) \sum_{i=1}^{N} (x_i - \bar{x})^2},
\end{equation}
where $x$ denotes the image matrix; $i$ and $j$ are patches in $x$; $\bar x$ is the mean of $x$; $N$ the the total number of units in $x$; and $w_{ij}$ is the spatial weight between pixels $i$ and $j$, for which we follow the common definition of $w_{ij} = 1$ if patches $i$ and $j$ are neighbors and 0 otherwise.


\begin{figure}[h!]
\includegraphics[width=\linewidth]{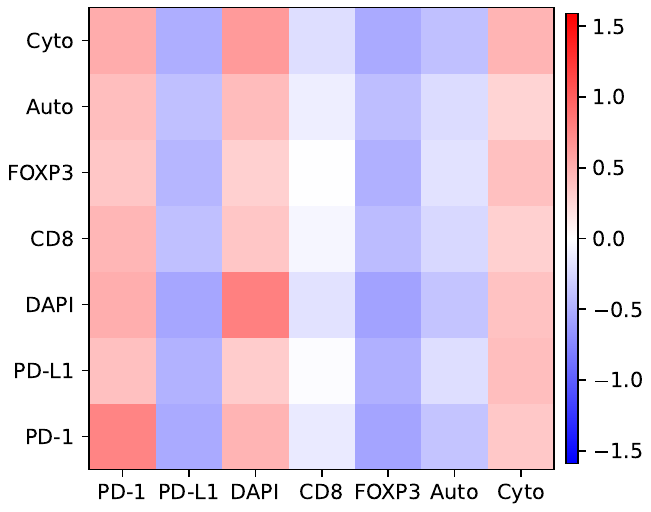}
\caption{Average marker attention matrix ($\mathbf A_k$) across the cohort for the Fluoroformer with ResNet embeddings. The displayed values are $z$-scored, meaning a value of $1$ indicates that the mean attention value for that entry is one standard deviation higher than the global mean across all markers. The $x$- and $y$-axes represent in- and outgoing attention, respectively.}
\label{fig4}
\end{figure}

\subsection{Training and implementation details}

All experiments were conducted on NVIDIA A100 GPUs with 80Gb of VRAM using the PyTorch \citep{paszke2019pytorch} software library. Lightning \citep{falcon2019pytorch} and Weights and Biases \citep{wandb} were employed to simplify training and logging. The AdamW optimization algorithm \citep{loshchilov2017decoupled} with a learning rate of $1 \times 10^{-4}$ was employed without a scheduler. All models were trained for a total of 25 epochs, which was sufficient for model convergence. The C-index was computed using the Lifelines package \citep{Davidson-Pilon2019} on the validation fold at the end of each training epoch, with model weights being checkpointed if a new maximum was reached. 

\section{Results}

\begin{table}[hbtp]
    \centering
    \caption{C-index and Moran's I (MI) by image type and embedding model}
    \label{tab:results-summary}
    \begin{tabular}{lcccc}
        \toprule
        \textbf{Image} & \textbf{Model} & \textbf{C-index $\pm$ STD} & \textbf{MI} \\
        \midrule
        \multirow{2}{*}{\textbf{mIF}} 
        & \textbf{ResNet} & $0.800 \pm 0.053$ & $0.501$\\
        & \textbf{UNI} & $0.807 \pm 0.056$ & $0.407$ \\
        \midrule
        \multirow{2}{*}{\textbf{H\&E}} 
        & \textbf{ResNet} & $0.771 \pm 0.043$ & $0.054$  \\
        & \textbf{UNI} & $0.825 \pm 0.077$ & $0.353$ \\
        \bottomrule
    \end{tabular}
\end{table}

\subsection{Prognostic performance}

The performance of the Fluoroformer approach is summarized in Table 1. Averaging across all 5 folds, the Fluoroformer achieves a C-index of 0.800 when using a ResNet50 embedder, compared to 0.771 for the H\&E-ABMIL baseline using the ResNet50 embedder. UNI improves H\&E performance as expected, with the Fluoroformer exhibiting a small increase in performance to 0.807. For comparison, the mIF-based CoxPH baseline based on commonly-used cell density metrics achieved a C-index of 0.689 $\pm$ 0.056. Thus, despite using off-the-shelf embedders optimized for H\&E and/or RGB images, the Fluoroformer strategy exhibits strong absolute and relative performance. 

\subsection{Marker-wise co-attention relationships}

A core benefit of the Fluoroformer approach is the generation of attention matrices ($\mathbf A_k$) between the different marker channels. We computed an average marker attention matrix to obtain an aggregated summary of the learned channel interactions. The average was computed by taking the 10\% most highly weighted patches for each WSI according to the vector $\mathbf a$, and is displayed in Figure~\ref{fig2} for ResNet50 and in the Appendix for UNI. Across both models, overall higher attention is observed towards the PD-1, DAPI, and cytokeratin channels.  


Moreover, we visualized spatial variation in marker attention across individual WSIs (Figure~\ref{fig3}). Specifically, in each matrix $\mathbf A_k$, we identified the marker receiving the most in-going attention by summing the entries in each column, then locating the index of the maximum value in the resulting 7-dimensional vector. We then visualized the resulting ``channel argmax heatmaps'' for each slide (Figure~\ref{fig3} for ResNet, Appendix Figure~\ref{fig7} for UNI). As expected based on the analysis above, PD-1 was commonly the most attended to channel. Cytokeratin also received the highest attention in regions of the tumor mass (Figure~\ref{fig3}AC), while DAPI received the most in alveolar structures (Figure~\ref{fig3}AB). Patches high in attention to PD-L1 were also observed, commonly directly adjacent to patches of high attention to cytokeratin (Figure~\ref{fig3}BC), with high levels of attention to CD8 and FOXP3 in regions that were often adjacent to the tissue, potentially relating to immune infiltration (Figure~\ref{fig3}D).

\subsection{Patch-wise attention heatmaps}

While the marker attention matrices offer insights into how the channels are combined per patch, the patch attention heatmaps from ABMIL indicate how the patches are combined into a WSI-level representation. Figure~\ref{fig4} contains exemplar patch attention heatmaps for the Fluoroformer and H\&E models, with a higher resolution version also included as Figure~\ref{fig5} in the Appendix. As quantified by Moran's I, the mIF-based Fluoroformer models exhibit higher spatial autocorrelation (i.e., smoothness) on average across the dataset (0.501 and 0.407 for Fluoroformer with ResNet and UNI, respectively, compared to 0.054 and 0.353 for the H\&E models). There are also visible differences in the regions most attended to by the different models. In the representative examples shown for instance, the Fluoroformer model more highly attended to tumor margins (Figure~\ref{fig4}BC), whereas the H\&E attention maps were largely concentrated on the tumor mass, indicating possibly complementary prognostic features. 




\section{Discussion and Conclusions}
In this work, we develop the Fluoroformer, a Transformer-inspired neural network architecture emphasizing biological interpretability and designed to scale attention-based multiple instance learning to multiplexed images. Using a dataset of 7-channel mIF WSIs from 416 NSCLC patients, the approach demonstrates strong performance in predicting patient prognosis. Importantly, the approach is flexible in terms of number of channels and embedders, even demonstrating similar performance to a H\&E-based model when using a H\&E foundational model. We expect even higher performance with mIF-optimized embedders in the future, though the variability in mIF assays presents challenges for the universality of such embedders. As such, we focused on developing a MIL strategy that can be applied to existing embedders. Beyond predictive performance, a key motivation for the strategy is its marker-wise attention interpretability. We highlight its potential for investigating spatial patterns in the tumor immune microenvironment, which is increasingly important in the age of immunotherapies. This interpretability is enhanced by the patch-level attention heatmaps, for which we observe higher spatial smoothness than H\&E based models. Future important work will involve further quantification and assessment of the observed patterns and their biological significance. As multiplexed, spatial biology techniques are increasingly used, the Fluoroformer may therefore serve as a general purpose method towards maximizing the utility of these rich data.

\acks{We thank Scott Rodig, James Lindsay, Jennifer Altreuter, and Katharina Hoebel for fruitful discussions and their support regarding the ImmunoProfile dataset. WL acknowledges support from the Ellison Foundation and the Wong Family Award in Translational Oncology.}

\bibliography{refs.bib}

\begin{thebibliography}{40}
\providecommand{\natexlab}[1]{#1}
\providecommand{\url}[1]{\texttt{#1}}
\expandafter\ifx\csname urlstyle\endcsname\relax
  \providecommand{\doi}[1]{doi: #1}\else
  \providecommand{\doi}{doi: \begingroup \urlstyle{rm}\Url}\fi

\bibitem[Ba et~al.(2016)Ba, Kiros, and Hinton]{ba2016layer}
Jimmy~Lei Ba, Jamie~Ryan Kiros, and Geoffrey~E Hinton.
\newblock Layer normalization.
\newblock \emph{arXiv preprint arXiv:1607.06450}, 2016.

\bibitem[Biewald(2020)]{wandb}
Lukas Biewald.
\newblock Experiment tracking with weights and biases, 2020.
\newblock URL \url{https://www.wandb.com/}.
\newblock Software available from wandb.com.

\bibitem[Carbonneau et~al.(2018)Carbonneau, Cheplygina, Granger, and Gagnon]{carbonneau2018multiple}
Marc-Andr{\'e} Carbonneau, Veronika Cheplygina, Eric Granger, and Ghyslain Gagnon.
\newblock Multiple instance learning: A survey of problem characteristics and applications.
\newblock \emph{Pattern Recognition}, 77:\penalty0 329--353, 2018.

\bibitem[Chen et~al.(2021)Chen, Lu, Weng, Chen, Williamson, Manz, Shady, and Mahmood]{chen2021multimodal}
Richard~J Chen, Ming~Y Lu, Wei-Hung Weng, Tiffany~Y Chen, Drew~FK Williamson, Trevor Manz, Maha Shady, and Faisal Mahmood.
\newblock Multimodal co-attention transformer for survival prediction in gigapixel whole slide images.
\newblock In \emph{Proceedings of the IEEE/CVF international conference on computer vision}, pages 4015--4025, 2021.

\bibitem[Chen et~al.(2024)Chen, Ding, Lu, Williamson, Jaume, Song, Chen, Zhang, Shao, Shaban, et~al.]{chen2024towards}
Richard~J Chen, Tong Ding, Ming~Y Lu, Drew~FK Williamson, Guillaume Jaume, Andrew~H Song, Bowen Chen, Andrew Zhang, Daniel Shao, Muhammad Shaban, et~al.
\newblock Towards a general-purpose foundation model for computational pathology.
\newblock \emph{Nature Medicine}, 30\penalty0 (3):\penalty0 850--862, 2024.

\bibitem[Cox(1972)]{cox1972regression}
David~R Cox.
\newblock Regression models and life-tables.
\newblock \emph{Journal of the Royal Statistical Society: Series B (Methodological)}, 34\penalty0 (2):\penalty0 187--202, 1972.

\bibitem[Davidson-Pilon(2019)]{Davidson-Pilon2019}
Cameron Davidson-Pilon.
\newblock lifelines: survival analysis in python.
\newblock \emph{Journal of Open Source Software}, 4\penalty0 (40):\penalty0 1317, 2019.
\newblock \doi{10.21105/joss.01317}.
\newblock URL \url{https://doi.org/10.21105/joss.01317}.

\bibitem[Dosovitskiy et~al.(2020)Dosovitskiy, Beyer, Kolesnikov, Weissenborn, Zhai, Unterthiner, Dehghani, Minderer, Heigold, Gelly, et~al.]{dosovitskiy2020image}
Alexey Dosovitskiy, Lucas Beyer, Alexander Kolesnikov, Dirk Weissenborn, Xiaohua Zhai, Thomas Unterthiner, Mostafa Dehghani, Matthias Minderer, Georg Heigold, Sylvain Gelly, et~al.
\newblock An image is worth 16x16 words: Transformers for image recognition at scale.
\newblock \emph{arXiv preprint arXiv:2010.11929}, 2020.

\bibitem[Falcon(2019)]{falcon2019pytorch}
William Falcon.
\newblock Pytorch lightning.
\newblock \emph{GitHub. Note: https://github.com/PyTorchLightning/pytorch-lightning}, 2019.

\bibitem[Ghahremani et~al.(2022)Ghahremani, Li, Kaufman, Vanguri, Greenwald, Angelo, Hollmann, and Nadeem]{ghahremani2022deep}
Parmida Ghahremani, Yanyun Li, Arie Kaufman, Rami Vanguri, Noah Greenwald, Michael Angelo, Travis~J Hollmann, and Saad Nadeem.
\newblock Deep learning-inferred multiplex immunofluorescence for immunohistochemical image quantification.
\newblock \emph{Nature machine intelligence}, 4\penalty0 (4):\penalty0 401--412, 2022.

\bibitem[Han et~al.(2020)Han, Liu, and Li]{han2020pd}
Yanyan Han, Dandan Liu, and Lianhong Li.
\newblock Pd-1/pd-l1 pathway: current researches in cancer.
\newblock \emph{American journal of cancer research}, 10\penalty0 (3):\penalty0 727, 2020.

\bibitem[He et~al.(2016)He, Zhang, Ren, and Sun]{he2016deep}
Kaiming He, Xiangyu Zhang, Shaoqing Ren, and Jian Sun.
\newblock Deep residual learning for image recognition.
\newblock In \emph{Proceedings of the IEEE conference on computer vision and pattern recognition}, pages 770--778, 2016.

\bibitem[Hendrycks and Gimpel(2016)]{hendrycks2016gaussian}
Dan Hendrycks and Kevin Gimpel.
\newblock Gaussian error linear units (gelus).
\newblock \emph{arXiv preprint arXiv:1606.08415}, 2016.

\bibitem[Hoebel et~al.(2024)Hoebel, Lindsay, Alessi, Weirather, Dryg, Altreuter, Awad, Rodig, and Lotter]{hoebel2024deep}
Katharina~Viktoria Hoebel, James~R Lindsay, Joao~V Alessi, Jason~L Weirather, Ian~D Dryg, Jennifer Altreuter, Mark~M Awad, Scott~J Rodig, and William~E Lotter.
\newblock Deep-learning model trained on multiplex immunofluorescence-stained tissue samples predicts the survival of patients with non-small cell lung cancer better than pd-l1 tps alone.
\newblock \emph{Cancer Research}, 84\penalty0 (6\_Supplement):\penalty0 6189--6189, 2024.

\bibitem[Ilse et~al.(2018)Ilse, Tomczak, and Welling]{ilse2018attention}
Maximilian Ilse, Jakub Tomczak, and Max Welling.
\newblock Attention-based deep multiple instance learning.
\newblock In \emph{International conference on machine learning}, pages 2127--2136. PMLR, 2018.

\bibitem[Javed et~al.(2022)Javed, Juyal, Padigela, Taylor-Weiner, Yu, and Prakash]{javed2022additive}
Syed~Ashar Javed, Dinkar Juyal, Harshith Padigela, Amaro Taylor-Weiner, Limin Yu, and Aaditya Prakash.
\newblock Additive mil: Intrinsically interpretable multiple instance learning for pathology.
\newblock \emph{Advances in Neural Information Processing Systems}, 35:\penalty0 20689--20702, 2022.

\bibitem[Katzman et~al.(2018)Katzman, Shaham, Cloninger, Bates, Jiang, and Kluger]{katzman2018deepsurv}
Jared~L Katzman, Uri Shaham, Alexander Cloninger, Jonathan Bates, Tingting Jiang, and Yuval Kluger.
\newblock Deepsurv: personalized treatment recommender system using a cox proportional hazards deep neural network.
\newblock \emph{BMC medical research methodology}, 18:\penalty0 1--12, 2018.

\bibitem[Lee et~al.(2020)Lee, Ren, Marella, Wang, Hartke, and Couto]{lee2020multiplex}
Chung-Wein Lee, Yan~J Ren, Mathieu Marella, Maria Wang, James Hartke, and Suzana~S Couto.
\newblock Multiplex immunofluorescence staining and image analysis assay for diffuse large b cell lymphoma.
\newblock \emph{Journal of immunological methods}, 478:\penalty0 112714, 2020.

\bibitem[Lindsay et~al.(2023)Lindsay, Sharma, Felt, Giobbie-Hurder, Dryg, Weirather, Altreuter, Mazor, Kumari, Alessi, et~al.]{lindsay2023immunoprofile}
James Lindsay, Bijaya Sharma, Kristen~D Felt, Anita Giobbie-Hurder, Ian Dryg, Jason~L Weirather, Jennifer Altreuter, Tali Mazor, Priti Kumari, Joao~V Alessi, et~al.
\newblock Immunoprofile: A prospective implementation of clinically validated, quantitative immune cell profiling test identifies tumor-infiltrating cd8+ and pd-1+ cell densities as prognostic biomarkers across a 2,023 patient pan-cancer cohort treated with different therapies.
\newblock \emph{Cancer Research}, 83\penalty0 (7\_Supplement):\penalty0 5706--5706, 2023.

\bibitem[Loshchilov and Hutter(2017)]{loshchilov2017decoupled}
Ilya Loshchilov and Frank Hutter.
\newblock Decoupled weight decay regularization.
\newblock \emph{arXiv preprint arXiv:1711.05101}, 2017.

\bibitem[Lu et~al.(2021)Lu, Williamson, Chen, Chen, Barbieri, and Mahmood]{lu2021data}
Ming~Y Lu, Drew~FK Williamson, Tiffany~Y Chen, Richard~J Chen, Matteo Barbieri, and Faisal Mahmood.
\newblock Data-efficient and weakly supervised computational pathology on whole-slide images.
\newblock \emph{Nature biomedical engineering}, 5\penalty0 (6):\penalty0 555--570, 2021.

\bibitem[Maron and Lozano-P{\'e}rez(1997)]{maron1997framework}
Oded Maron and Tom{\'a}s Lozano-P{\'e}rez.
\newblock A framework for multiple-instance learning.
\newblock \emph{Advances in neural information processing systems}, 10, 1997.

\bibitem[Moran(1950)]{moran1950notes}
Patrick~AP Moran.
\newblock Notes on continuous stochastic phenomena.
\newblock \emph{Biometrika}, 37\penalty0 (1/2):\penalty0 17--23, 1950.

\bibitem[Mu{\~n}oz-Castro et~al.(2022)Mu{\~n}oz-Castro, Noori, Magdamo, Li, Marks, Frosch, Das, Hyman, and Serrano-Pozo]{munoz2022cyclic}
Clara Mu{\~n}oz-Castro, Ayush Noori, Colin~G Magdamo, Zhaozhi Li, Jordan~D Marks, Matthew~P Frosch, Sudeshna Das, Bradley~T Hyman, and Alberto Serrano-Pozo.
\newblock Cyclic multiplex fluorescent immunohistochemistry and machine learning reveal distinct states of astrocytes and microglia in normal aging and alzheimer’s disease.
\newblock \emph{Journal of Neuroinflammation}, 19\penalty0 (1):\penalty0 30, 2022.

\bibitem[Oquab et~al.(2023)Oquab, Darcet, Moutakanni, Vo, Szafraniec, Khalidov, Fernandez, Haziza, Massa, El-Nouby, et~al.]{oquab2023dinov2}
Maxime Oquab, Timoth{\'e}e Darcet, Th{\'e}o Moutakanni, Huy Vo, Marc Szafraniec, Vasil Khalidov, Pierre Fernandez, Daniel Haziza, Francisco Massa, Alaaeldin El-Nouby, et~al.
\newblock Dinov2: Learning robust visual features without supervision.
\newblock \emph{arXiv preprint arXiv:2304.07193}, 2023.

\bibitem[Otsu et~al.(1975)]{otsu1975threshold}
Nobuyuki Otsu et~al.
\newblock A threshold selection method from gray-level histograms.
\newblock \emph{Automatica}, 11\penalty0 (285-296):\penalty0 23--27, 1975.

\bibitem[Otter et~al.(2020)Otter, Medina, and Kalita]{otter2020survey}
Daniel~W Otter, Julian~R Medina, and Jugal~K Kalita.
\newblock A survey of the usages of deep learning for natural language processing.
\newblock \emph{IEEE transactions on neural networks and learning systems}, 32\penalty0 (2):\penalty0 604--624, 2020.

\bibitem[Paszke et~al.(2019)Paszke, Gross, Massa, Lerer, Bradbury, Chanan, Killeen, Lin, Gimelshein, Antiga, et~al.]{paszke2019pytorch}
Adam Paszke, Sam Gross, Francisco Massa, Adam Lerer, James Bradbury, Gregory Chanan, Trevor Killeen, Zeming Lin, Natalia Gimelshein, Luca Antiga, et~al.
\newblock Pytorch: An imperative style, high-performance deep learning library.
\newblock \emph{Advances in neural information processing systems}, 32, 2019.

\bibitem[Peng et~al.(2023)Peng, Wu, Liu, He, Xie, Zhong, Liu, Tang, Li, Xiong, et~al.]{peng2023multiplex}
Haoxin Peng, Xiangrong Wu, Shaopeng Liu, Miao He, Chao Xie, Ran Zhong, Jun Liu, Chenshuo Tang, Caichen Li, Shan Xiong, et~al.
\newblock Multiplex immunofluorescence and single-cell transcriptomic profiling reveal the spatial cell interaction networks in the non-small cell lung cancer microenvironment.
\newblock \emph{Clinical and Translational Medicine}, 13\penalty0 (1):\penalty0 e1155, 2023.

\bibitem[Raskov et~al.(2021)Raskov, Orhan, Christensen, and Gögenur]{raskov2021}
Hans Raskov, Adile Orhan, Jan~Pravsgaard Christensen, and Ismail Gögenur.
\newblock Cytotoxic cd8+ t cells in cancer and cancer immunotherapy.
\newblock \emph{British Journal of Cancer}, 124:\penalty0 359–367, 2021.

\bibitem[Rudensky(2011)]{Rudensky2011}
Alexander~Y. Rudensky.
\newblock Regulatory t cells and foxp3.
\newblock \emph{Immunological Reviews}, 241\penalty0 (1):\penalty0 260--268, 2011.
\newblock \doi{https://doi.org/10.1111/j.1600-065X.2011.01018.x}.
\newblock URL \url{https://onlinelibrary.wiley.com/doi/abs/10.1111/j.1600-065X.2011.01018.x}.

\bibitem[Shao et~al.(2021)Shao, Bian, Chen, Wang, Zhang, Ji, et~al.]{shao2021transmil}
Zhuchen Shao, Hao Bian, Yang Chen, Yifeng Wang, Jian Zhang, Xiangyang Ji, et~al.
\newblock Transmil: Transformer based correlated multiple instance learning for whole slide image classification.
\newblock \emph{Advances in neural information processing systems}, 34:\penalty0 2136--2147, 2021.

\bibitem[Sorin et~al.(2023)Sorin, Karimi, Rezanejad, Miranda, Desharnais, McDowell, Dor{\'e}, Arabzadeh, Breton, Fiset, et~al.]{sorin2023single}
Mark Sorin, Elham Karimi, Morteza Rezanejad, W~Yu Miranda, Lysanne Desharnais, Sheri~AC McDowell, Samuel Dor{\'e}, Azadeh Arabzadeh, Valerie Breton, Benoit Fiset, et~al.
\newblock Single-cell spatial landscape of immunotherapy response reveals mechanisms of cxcl13 enhanced antitumor immunity.
\newblock \emph{Journal for Immunotherapy of Cancer}, 11\penalty0 (2), 2023.

\bibitem[Sudharshan et~al.(2019)Sudharshan, Petitjean, Spanhol, Oliveira, Heutte, and Honeine]{sudharshan2019multiple}
PJ~Sudharshan, Caroline Petitjean, Fabio Spanhol, Luiz~Eduardo Oliveira, Laurent Heutte, and Paul Honeine.
\newblock Multiple instance learning for histopathological breast cancer image classification.
\newblock \emph{Expert Systems with Applications}, 117:\penalty0 103--111, 2019.

\bibitem[Vaswani(2017)]{vaswani2017attention}
Ashish Vaswani.
\newblock Attention is all you need.
\newblock \emph{arXiv preprint arXiv:1706.03762}, 2017.

\bibitem[Wilson et~al.(2021)Wilson, Ospina, Townsend, Nguyen, Moran~Segura, Schildkraut, Tworoger, Peres, and Fridley]{wilson2021challenges}
Christopher~M Wilson, Oscar~E Ospina, Mary~K Townsend, Jonathan Nguyen, Carlos Moran~Segura, Joellen~M Schildkraut, Shelley~S Tworoger, Lauren~C Peres, and Brooke~L Fridley.
\newblock Challenges and opportunities in the statistical analysis of multiplex immunofluorescence data.
\newblock \emph{Cancers}, 13\penalty0 (12):\penalty0 3031, 2021.

\bibitem[Wu et~al.(2022)Wu, Trevino, Wu, Swanson, Kim, D’Angio, Preska, Charville, Dalerba, Egloff, et~al.]{wu2022graph}
Zhenqin Wu, Alexandro~E Trevino, Eric Wu, Kyle Swanson, Honesty~J Kim, H~Blaize D’Angio, Ryan Preska, Gregory~W Charville, Piero~D Dalerba, Ann~Marie Egloff, et~al.
\newblock Graph deep learning for the characterization of tumour microenvironments from spatial protein profiles in tissue specimens.
\newblock \emph{Nature Biomedical Engineering}, 6\penalty0 (12):\penalty0 1435--1448, 2022.

\bibitem[Xu et~al.(2024)Xu, Usuyama, Bagga, Zhang, Rao, Naumann, Wong, Gero, Gonz{\'a}lez, Gu, et~al.]{xu2024whole}
Hanwen Xu, Naoto Usuyama, Jaspreet Bagga, Sheng Zhang, Rajesh Rao, Tristan Naumann, Cliff Wong, Zelalem Gero, Javier Gonz{\'a}lez, Yu~Gu, et~al.
\newblock A whole-slide foundation model for digital pathology from real-world data.
\newblock \emph{Nature}, pages 1--8, 2024.

\bibitem[Zadeh and Schmid(2020)]{zadeh2020bias}
Shekoufeh~Gorgi Zadeh and Matthias Schmid.
\newblock Bias in cross-entropy-based training of deep survival networks.
\newblock \emph{IEEE transactions on pattern analysis and machine intelligence}, 43\penalty0 (9):\penalty0 3126--3137, 2020.

\bibitem[Zhou(2018)]{zhou2018brief}
Zhi-Hua Zhou.
\newblock A brief introduction to weakly supervised learning.
\newblock \emph{National science review}, 5\penalty0 (1):\penalty0 44--53, 2018.

\end{thebibliography}

\appendix

\section*{Appendix}
Appendix figures are displayed in subsequent pages.

\begin{figure*}[h!]
\centering
\includegraphics[width=\textwidth]{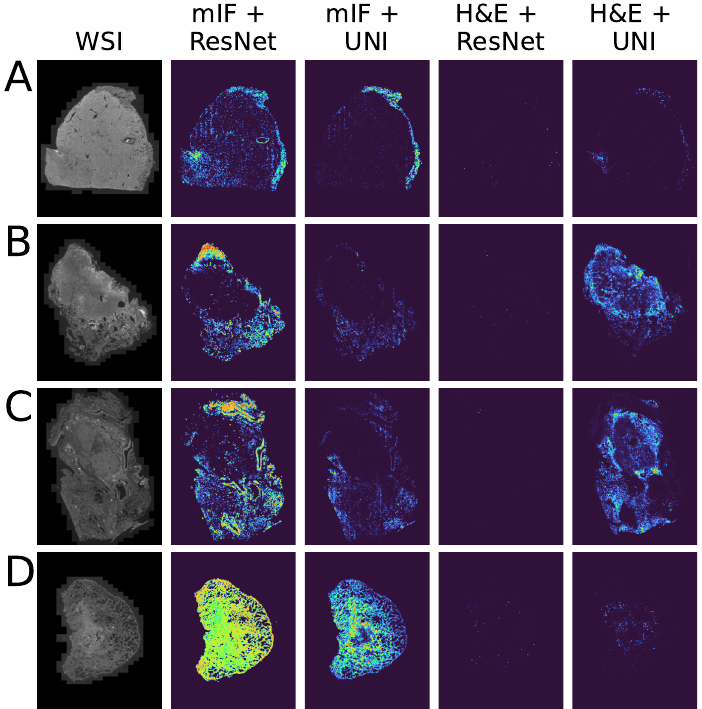}
\caption{Heatmaps of patch attention for each combination of imaging modality and neural embedder. The data is the same as displayed in the main text, but at high resolution.}
\label{fig5}
\end{figure*}

\begin{figure*}[t!]
\centering
\includegraphics[width=\linewidth]{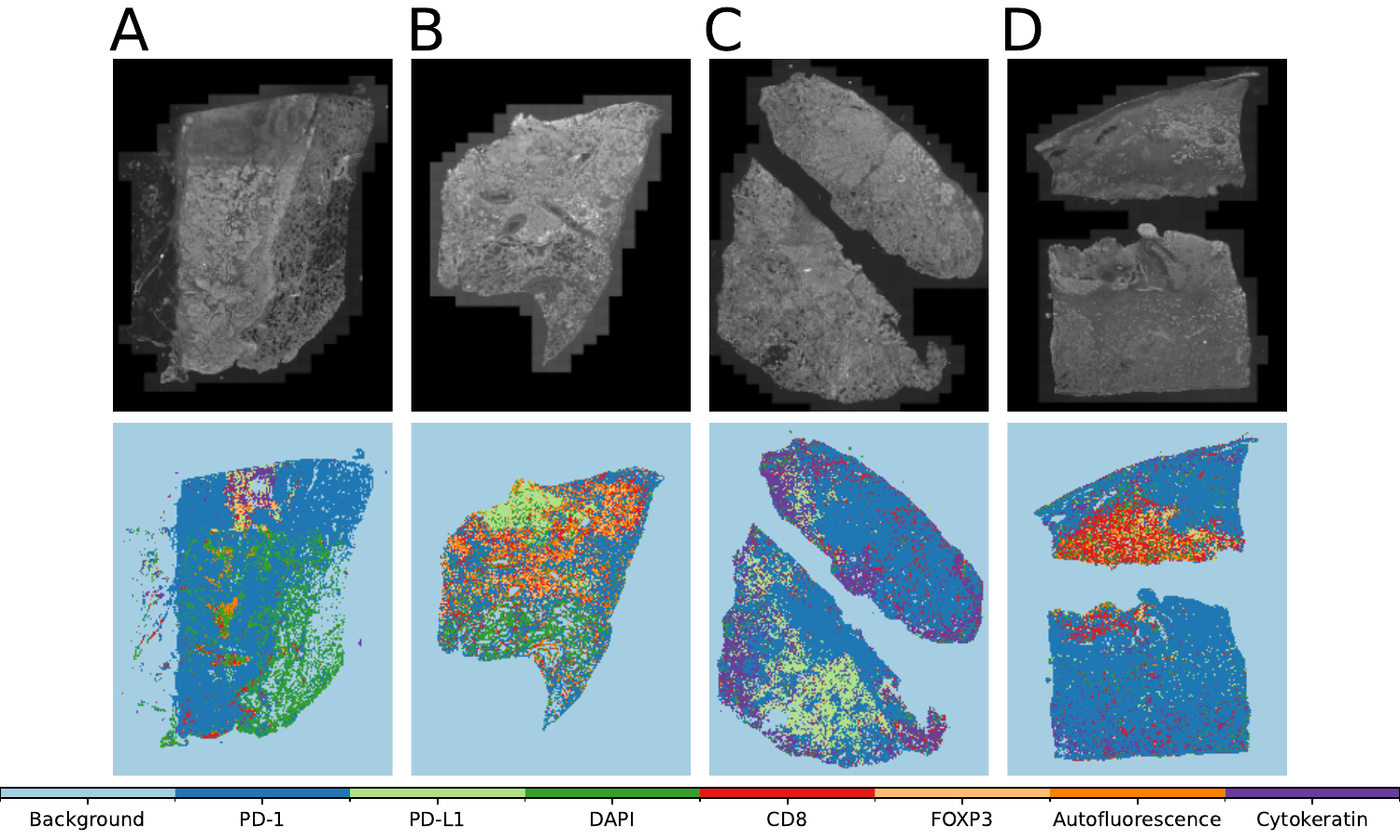}
\caption{Heatmaps of the most highly attended markers in each patch of selected mIF images for the Fluoroformer model with a UNI embedder.}
\label{fig6}
\end{figure*}

\begin{figure*}[h!]
\centering
\includegraphics[width=0.75\linewidth]{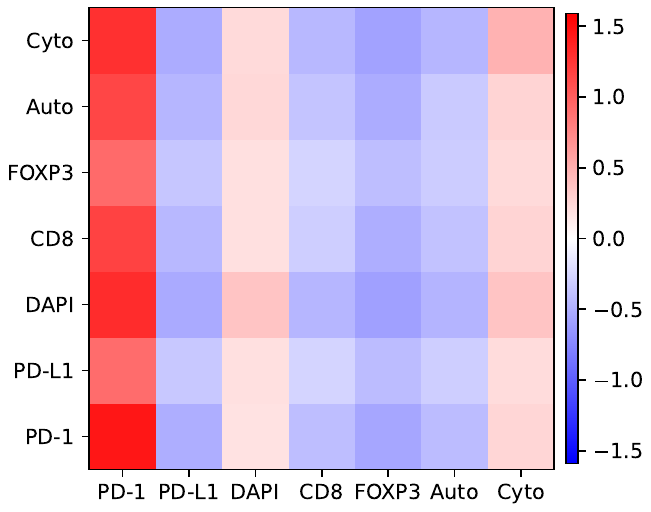}
\caption{Average marker attention matrix ($\mathbf A_k$) across the cohort for the Fluoroformer with UNI embeddings. The displayed values are $z$-scored, meaning a value of $1$ indicates that the mean attention value for that entry is one standard deviation higher than the global mean across all markers. The $x$- and $y$-axes represent in- and outgoing attention, respectively.}
\label{fig7}
\end{figure*}

\end{document}